\documentclass[a4paper,11pt]{article}
\usepackage{float}
\usepackage{pos}
\usepackage{amsmath,bm,slashed}
\usepackage{amsthm}
\usepackage{xspace}
\usepackage[capitalize]{cleveref}

\newcommand{\Vcb}[0]{|V_{\rm cb}|}
\newcommand{\Vub}[0]{|V_{\rm ub}|}

\newcommand{\gvar}{\textsc{gvar}\xspace}
\newcommand{\lsqfit}{\textsc{lsqfit}\xspace}

\newcommand{\tmin}{t_{\rm min}\xspace}

\newcommand{\vacuum}{\emptyset}

\newcommand{\matrixel}[3]{\left< #1 \vphantom{#2#3} \right| #2 \left| #3 \vphantom{#1#2} \right>} % for Dirac matrix elements
\DeclareMathOperator{\arccosh}{arcCosh}

\newcommand{\finalstate}{P}
\newcommand{\Bpi}{\ensuremath{B\to\pi}}

\title{$B\to\pi$, $B_{(s)}\to D_{(s)}$ from 2+1+1 Flavor Lattice QCD}

\author*[a]{Nicholas Cassar}
\author*[b,c]{Akhil Chauhan}
\author[d]{Carleton DeTar}
\author[b,c]{Aida El-Khadra}
\author[e]{Elvira G\'amiz}
\author[f]{Steven Gottlieb}
\author[a]{William Jay}
\author[g]{Andreas S. Kronfeld}
\author[h]{Jack Laiho}
\author[b,c]{Andrew Lytle}
\author[i]{Alejandro Vaquero}

\affiliation[a]{Department of Physics, Colorado State University,\\
  Fort Collins, Colorado 80523, USA}

\affiliation[b]{Department of Physics, University of Illinois Urbana-Champaign,\\
Urbana, Illinois 61801, USA}

\affiliation[c]{Illinois Center for Advanced Studies of the Universe, University of Illinois, \\ Urbana, IL 61801, United States}

\affiliation[d] {Department of Physics and Astronomy, University of Utah,\\ Salt Lake City, Utah 84112, USA}

\affiliation[e]{Departamento de Física Teórica y del Cosmos, Universidad de Granada,\\ E-18071, Granada,
Spain}

\affiliation[f]{Department of Physics, Indiana University,\\ Bloomington, Indiana 47405, USA}

\affiliation[g]{Theory Division, Fermi National Accelerator Laboratory,\\ Batavia, Illinois 60510, USA}

\affiliation[h]{Department of Physics, Syracuse University,\\ Syracuse, New York 13244, USA}

\affiliation[i]{Departmento de F\'isica Te\'orica, Universidad de Zaragoza and CAPA,\\ CP50009, Zaragoza, Spain}

\emailAdd{nick.cassar@colostate.edu}
\emailAdd{akhilc2@illinois.edu}

\abstract{
\vspace*{-2mm}
\textbf{\textsf{Fermilab Lattice and MILC Collaborations}}\\[0.7em]
We present a lattice-QCD calculation of the hadronic form factors for the semileptonic decays $B \to \pi$ and $B_{(s)} \to D_{(s)} \ell \nu$, computed using the highly improved staggered quark action for both valence and sea quarks on the MILC Collaboration’s $2+1+1$-flavor ensembles with lattice spacings ranging from $0.09$~fm to $0.03$~fm, many with physical pion masses.
On our finest ensembles, we compute the form factors directly at the physical $b$-quark mass. 
We discuss the computational setup and analysis strategies for two- and three-point correlation functions. For $B_{(s)} \to D_{(s)}$, we present preliminary results of chiral-continuum fits for the scalar and vector form factors. 
The goal of this project is a percent-level determination of the scalar and vector form factors to enable high-precision determinations of $\Vub$ and $\Vcb$.
This work fits into a broader program of lattice-QCD studies of weak $B$-meson decays by the Fermilab Lattice and MILC Collaborations.
}

\FullConference{
}

\begin{document}
\maketitle

\section{Introduction}
\begin{figure}[b] 
    \centering
    \includegraphics[width=0.49\linewidth]{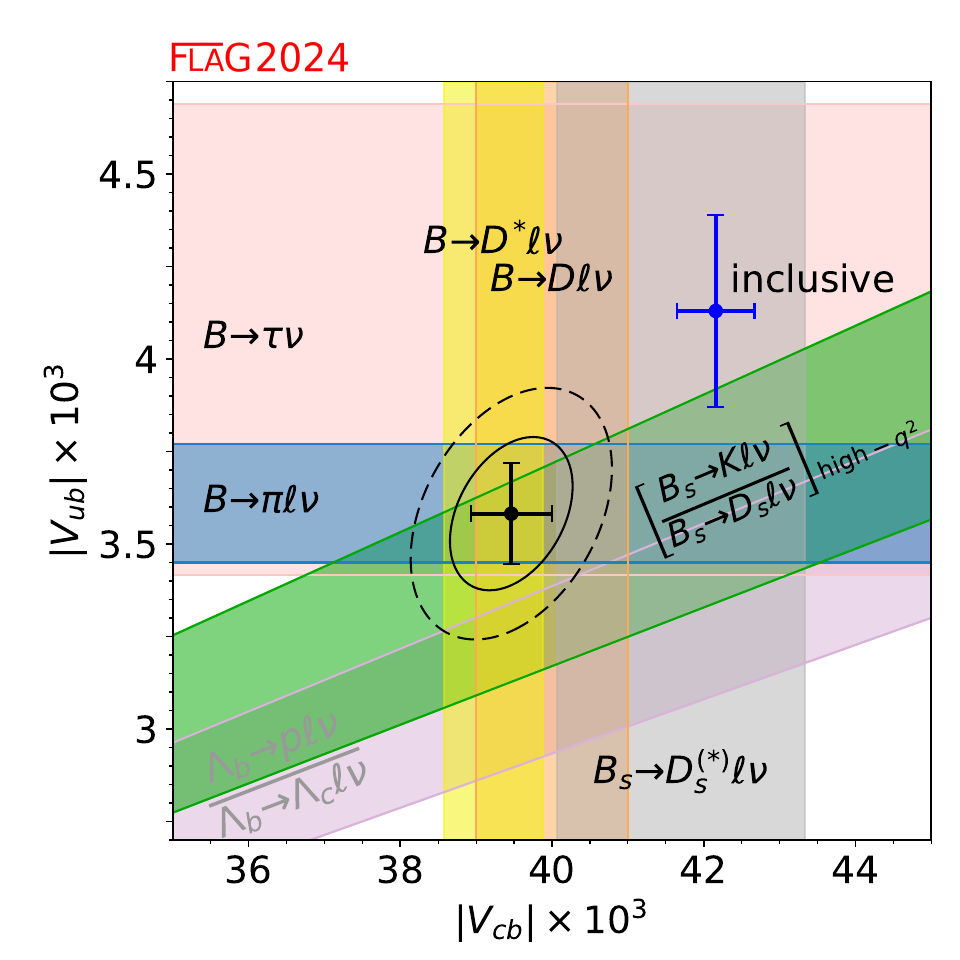}
    \includegraphics[width=0.49\linewidth]{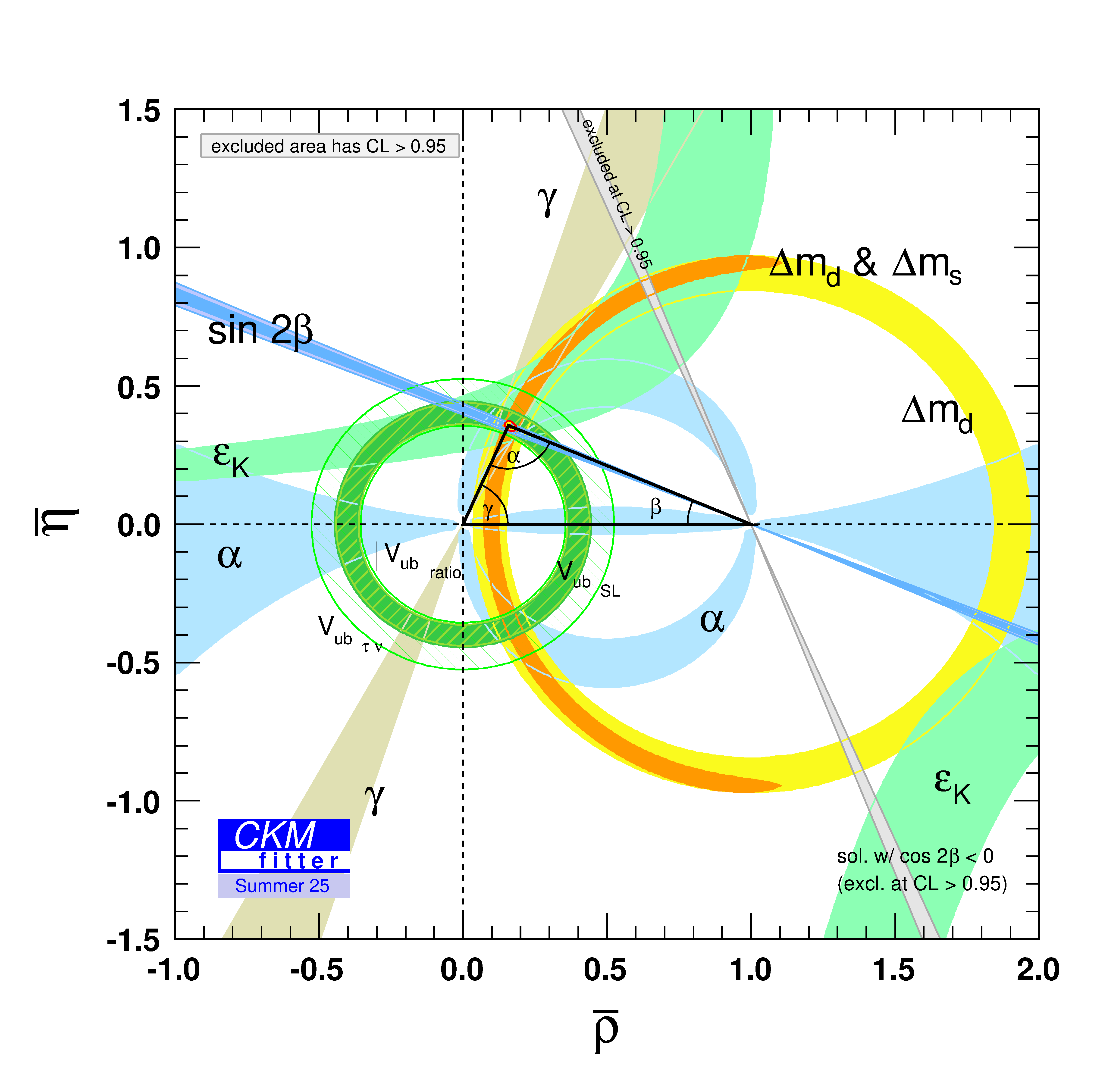}
    \caption{\textbf{Left:} Summary of tensions between inclusive and exclusive determinations of $\Vcb$ and $\Vub$, reproduced from \cite{FlavourLatticeAveragingGroupFLAG:2024oxs}.
    A primary goal of the present work is to reduce the theoretical uncertainty for the exclusive decays $B\to\pi\ell\nu$ and $B_{(s)}\to D_{(s)}\ell \nu$ to the roughly $1\%$ level, in line with near-term experimental goals.
    \textbf{Right:}
    Global-fit results from the \textsc{CKMfitter} group~\cite{Charles:2004jd} highlighting constraints on the parameters $\alpha,\beta,\gamma$ of the CKM triangle. 
    The parameter $\alpha$ associated with the apex of the CKM triangle is proportional to the ratio $\Vub/\Vcb$. 
    }
    \label{fig:flag_vub_vcb}
\end{figure}

Semileptonic $B_{(s)}$ decays provide an avenue for high-precision determinations of CKM matrix elements as well as a probe for potential new physics.
Hadronic form factors---non-perturbative quantities defined within quantum chromodynamics (QCD)---are essential ingredients in the Standard-Model prediction for exclusive decay rates.
This work describes ongoing work to compute the QCD contributions for three exclusive semileptonic decays: $B \to \pi\ell\nu$ and $B_{(s)} \to D_{(s)}\ell\nu$, allowing for the determination of $\Vub$, $\Vcb$ and lepton flavor universality (LFU) ratios $R(D_{(s)})$.
The current status of $\Vub$ and $\Vcb$ extracted from the most recent FLAG report \cite{FlavourLatticeAveragingGroupFLAG:2024oxs} is shown in the LHS of \Cref{fig:flag_vub_vcb}.  On the RHS, constraints on the CKM triangle from the CKMFitter group \cite{Charles:2004jd} are displayed.
There remains a long-standing discrepancy between inclusive and exclusive determinations of these CKM matrix elements.
This proceedings reports on ongoing work aimed at providing more precise predictions of the relevant hadronic form factors---with improved systematic control---to help understand the source of this tension.

Currently $\Vub$ and $\Vcb$ have uncertainties of ~1.3 and 3.9\% respectively \cite{FlavourLatticeAveragingGroupFLAG:2024oxs}. Ongoing experimental efforts from Belle II and LHCb aim to measure new hadronic observables and decay rates with percent-level precision. New theory calculations of $B$ meson observables with commensurate precisions are required to shed light on these $B$-anomalies. 

In these proceedings, we present an update from the Fermilab Lattice and MILC Collaborations ongoing work to calculate the form factors for the exclusive semileptonic decay $B \rightarrow \pi$, $B_{(s)} \to D_{(s)}$ via lattice QCD using highly-improved staggered quarks (HISQ), both light and heavy, in both the valence and sea. All three channels use the same analysis strategy.
We first focus on $B \to \pi$ for the analysis of the two- and three-point function, while $B_{(s)} \to D_{(s)}$  decays focus on the continuum fit forms and error budget due to the statistical advantage of having a strange quark spectator.

\section{Calculation details \label{sec:calculation_details}}
\begin{figure}
    \centering
    \includegraphics[width=0.75\linewidth]{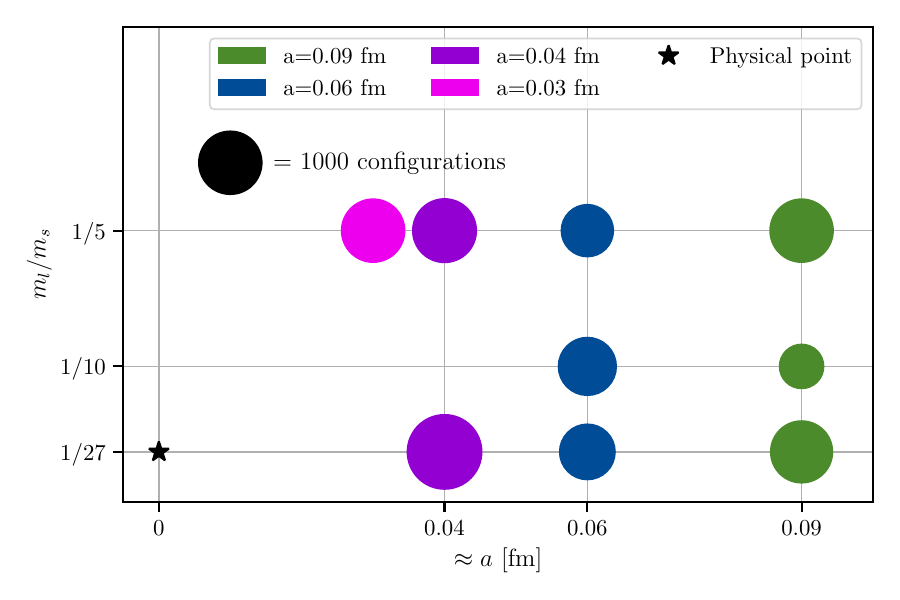}
    \caption{Summary of lattice spacings and light quark masses used in all decay channels. The area of the circles correspond to the number of configurations in the ensemble}
    \label{fig:ensembles}
\end{figure}

Our last update \cite{Lytle:2024zfr} presented preliminary results based on lattice spacings ranging from $a=0.09$~fm down to $0.04$~fm, and pion masses ranging from $\sim$135--330 MeV. Here we include preliminary results from an ensemble with a lattice spacing of $a\approx 0.03$~fm, as well as an ensemble at $a\approx 0.04$~fm tuned to the physical pion mass, as shown in \cref{fig:ensembles}. The new ensembles anchor the continuum extrapolation.
All ensembles have been generated by the MILC Collaboration~\cite{MILC:2010pul,MILC:2012znn,Bazavov:2017lyh} using $N_f =2+1+1$ sea quarks with the HISQ action. 

Two-point and three-point functions are also computed using the HISQ action for the valence quarks which gives improved control of discretization effects stemming from heavy quarks in $B$-meson decays.
We calculate all correlation functions for a set of heavy quark masses $m_h$ ranging from the charm mass up to $am_h \simeq 1$. For the finest lattice spacings we reach the $b$ quark mass $am_b$ without extrapolation. 

We blind our analysis of the lattice data: 
For each decay channel, the relevant three-point correlation functions are multiplied by a common blinding factor drawn randomly from the interval [0.95, 1.05].
The blinding factor will be removed after the analysis is finalized.

\section{Correlators, form factors, renormalization}

The analysis of correlation functions proceeds in two steps: First, we fit the two-point functions to obtain an estimate of the energies and amplitudes of the initial- and final-state hadrons. This information is then used as input to guide joint correlated fits of the two-point and three-point functions, which furnish the required form factors as a function of momentum transfer.

All correlation function data are fit to the following spectral decompositions:
\begin{align}
    C_{\finalstate}^\mathcal{O}(t, \bm{p})
    &= \sum_{n=0} (-1)^{n(t+1)} \frac{\left|\matrixel{\vacuum}{\mathcal{O}_{\finalstate}}{n}\right|^2}{2E_\finalstate^{(n)}(\bm{p})}
    \left( e^{-E_\finalstate^{(n)}(\bm{p})t} + e^{-E_\finalstate^{(n)}(\bm{p})(N_t-t)} \right), \label{eq:2pt_spectral_decomp_final}\\
    C_{H}^\mathcal{O}(t)
    &= \sum_{m=0} (-1)^{m(t+1)} \frac{\left|\matrixel{\vacuum}{\mathcal{O}_{H}}{m}\right|^2}{2M_H^{(m)}}
    \left( e^{-M_H^{(m)}t} + e^{-M_H^{(m)}(N_t-t)} \right), \label{eq:2pt_spectral_decomp_initial}\\    
    \begin{split}
    C_{H \to \finalstate}^J(t,T,\bm{p})
    &= \sum_{m,n}
        (-1)^{n(t+1)} (-1)^{m(T-t-1)}
        \frac{
            \matrixel{\vacuum}{\mathcal{O}_\finalstate}{n}
            \matrixel{n}{J}{m}            
            \matrixel{m}{\mathcal{O}_H}{\vacuum}}
            {4 E_\finalstate^{(n)}(\bm{p}) M_H^{(m)}}\\
        & \phantom{000}\times
        \left(e^{-E_\finalstate^{(n)}(\bm{p})t} + e^{-E_\finalstate^{(n)}(\bm{p})(N_t - t)} \right)
        \left(e^{-M_H^{(m)}(T-t)} + e^{-M_H^{(m)}(N_t-T+t)}\right)
        \label{eq:3pt_spectral_decomp},
    \end{split}
\end{align}
$H$ denotes a generic pseudoscalar initial state consisting of a heavy quark and a light antiquark.
$\finalstate$ denotes a generic pseudoscalar final state, which can be either a pion or $D_{(s)}$ meson.
$\mathcal{O}_p$ and $\mathcal{O}_H$ are operators with the flavor quantum numbers to couple to the initial- and final-state hadrons and with $J^P=0^-$.

Our calculation is set in the rest frame of the decaying meson with the final-state meson computed at several different momenta $\mathbf{p}$.
In this frame, the correlators depend on the initial-state-meson mass, $M_H$, and the final-state-meson energy, $E_P(\mathbf{p})$.
To describe flavor-changing weak interactions we use three operators $J$ in \cref{eq:3pt_spectral_decomp}:
the scalar $S$, the temporal component of the vector current $V^0$, and the spatial component of the vector current $V^i$.

\Cref{eq:2pt_spectral_decomp_final,eq:2pt_spectral_decomp_initial,eq:3pt_spectral_decomp} contain the usual opposite-parity contributions, proportional to $(-1)^T$ or $(-1)^t$, which are present in all staggered-fermion calculations.
These contributions are present in all pion correlators with non-zero momentum and in all heavy-meson correlators. 
Additional details relating to the staggered-quark structure of the operators are described in \cite{FermilabLattice:2022gku}.

The three-point functions give access to the transition matrix elements that appear in the definitions of the form factors.
In the rest frame of the decaying heavy meson, the form factors are defined via:
\begin{align}
f_\parallel(q^2)	&= Z_{V^0} \frac{\matrixel{\finalstate}{V^0}{H}}{\sqrt{2 M_H}}, \label{eq:f_parallel}\\
f_\perp(q^2)		&= Z_{V^i} \frac{1}{p^i} \frac{\matrixel{\finalstate}{V^i}{H}}{\sqrt{2 M_H}}, \label{eq:f_perp}\\
f_0(q^2)			&= Z_m Z_S \frac{m_h-m_x}{M_H^2 - M_\finalstate^2} \matrixel{\finalstate}{S}{H}. \label{eq:f_0}\\
f_+(q^2) &= \left( \frac{M_H - E_\finalstate}{\sqrt{2 M_H}} \right)
        \left( 1 - \frac{E_\finalstate^2 - M_\finalstate^2}{(M_H - E_\finalstate)^2}\right) f_\perp(q^2)
        + \left( \frac{M_H^2 - M_\finalstate^2}{M_H - E_\finalstate}\right) \frac{f_0(q^2)}{2 M_H},
    \label{eq:fplus_perp_0}
\end{align}
[No sum is implied in \cref{eq:f_perp}.] 
The valence quark masses are written as $m_h$ for the initial heavy ``bottom" quark and $m_x$ for the final light $(B\to\pi\ell\nu)$ or charm $(B_{(s)}\to D_{(s)}\ell\nu)$ quark.

The form factors describe effects from the non-perturbative hadronic structure of the weak-interaction vertex.
The Standard-Model prediction for the differential decay rate contains two independent form form factors, conventionally $f_0$ and $f_+$, defined in \cref{eq:f_0,eq:fplus_perp_0} respectively.
These form factors will be used in combination with experimental data to extract the CKM matrix elements $\Vub$ and $\Vcb$.

\Cref{eq:f_parallel,eq:f_perp,eq:f_0} contain renormalization factors which relate the bare, lattice-regulated matrix elements to their continuum quantities relevant for phenomenology.
The required renormalization factors can be extracted using partial conservation of vector current (PVCV). 
For the matrix elements at hand, PCVC reads
\begin{align}
    Z_{V^0}(M_H - E_\finalstate) \matrixel{\finalstate}{V^0}{H} + Z_{V^i} \bm{q}\cdot \matrixel{\finalstate}{\bm{V}}{H} = Z_m Z_S (m_h - m_x) \matrixel{\finalstate}{S}{H}.
    \label{eq:PCVC}
\end{align}
A convenient consequence of the HISQ action used in the present calculation is that  $Z_mZ_S = 1$~\cite{Karsten:1980wd,Smit:1987zh}.
\Cref{eq:PCVC} gives a linear relation between $Z_{V^0}$ and $Z_{V^i}$, valid for any momentum transfer.
As in \cite{FermilabLattice:2022gku}, values for the renormalization factors will be extracted from a fit to \cref{eq:PCVC}.

\section{Correlator analysis}
\label{sec:correlator_analysis}
\subsection{Two-point functions}

\begin{figure}
    \centering
    \includegraphics[width=1.0\linewidth]{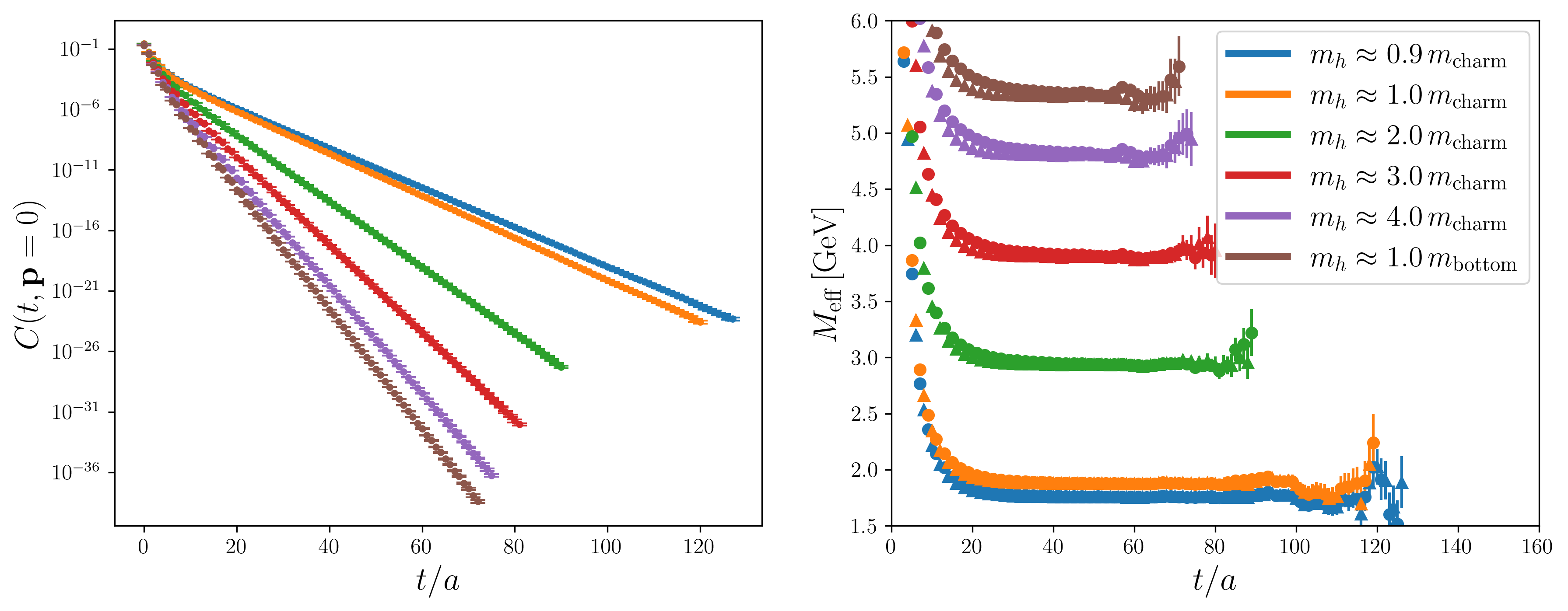}
    \caption{$B$-meson two-point functions on the physical-mass $a\approx0.04 \,\rm{fm}$ ensemble.
    \textbf{Left:}
    The correlation functions, excluding data with a noise-to-signal ratio exceeding $30\%$ that are not used in fits.
    The colors correspond to different heavy quark masses.
    \textbf{Right:}
    Effective masses for each correlator plotted separately on even and odd time sliices (denoted with triangles or circles), in approximate physical units.  Although noise increases with heavy quark mass, a large plateau region in Euclidean time is observed for all correlators.
    Note that that the largest heavy quark mass on this ensemble corresponds to a meson mass $M_B \approx 5.3 \,\rm{GeV}$.}
    \label{fig:bmeson2pt}
\end{figure}
\Cref{fig:bmeson2pt} shows data for a heavy-meson two-point function on the ensemble with $a\approx 0.04$~fm and with physical-mass pions.
Although reasonable plateaus are observed in the effective masses
$am_{\rm eff}(t) \equiv \frac{1}{2}\arccosh \left[ (C(t+2)+ C(t-2))/2C(t) \right]$,
the present analysis controls for possible excited-state contamination using Bayesian fits to \cref{eq:2pt_spectral_decomp_final,eq:2pt_spectral_decomp_initial} using \gvar and \lsqfit~\cite{gvar:2022,lsqfit,Lepage:2001ym}.
Correlator data with a noise-to-signal ratio of greater than $30\%$ are excluded from fits.
Bayesian priors for the energy levels are specified in terms of energy splittings, with
positivity of energy and amplitudes enforced in the usual way using log-normal priors. Central values for priors of the ground-state pion energy can be estimated from the effective mass $a M_{\rm eff}(t)$  or (on the physical-mass ensembles) from PDG value~\cite{ParticleDataGroup:2024cfk}; see \cite{FermilabLattice:2022gku} for more details.
Priors for final-state mesons at non-zero momentum are boosted from the ground state using the relativistic energy dispersion relation.

For heavy-meson two-point functions, where the ground-state mass ranges between $M_D$ and $M_B$, the central value for the prior is set using the effective mass.
For excited states, different methods were explored to set the central value for the prior.
For instance, one can use the observed spectrum of $B$-meson excited states with $J^P = 0^\pm$ to estimate expected regions of energy with a large spectral weight.
Alternatively, one can estimate the location of finite-volume energy levels based on the location of naive non-interacting multiparticle states with the desired quantum numbers.
Since we find that results for ground-state quantities are insensitive to the precise choice of excited-state priors, in the present analysis we choose the latter method, given the relative paucity of experimental results for the spectrum of excited $B$-mesons with $J^P=0^\pm$.

\begin{figure}
    \centering
    \includegraphics[width=0.75\linewidth]{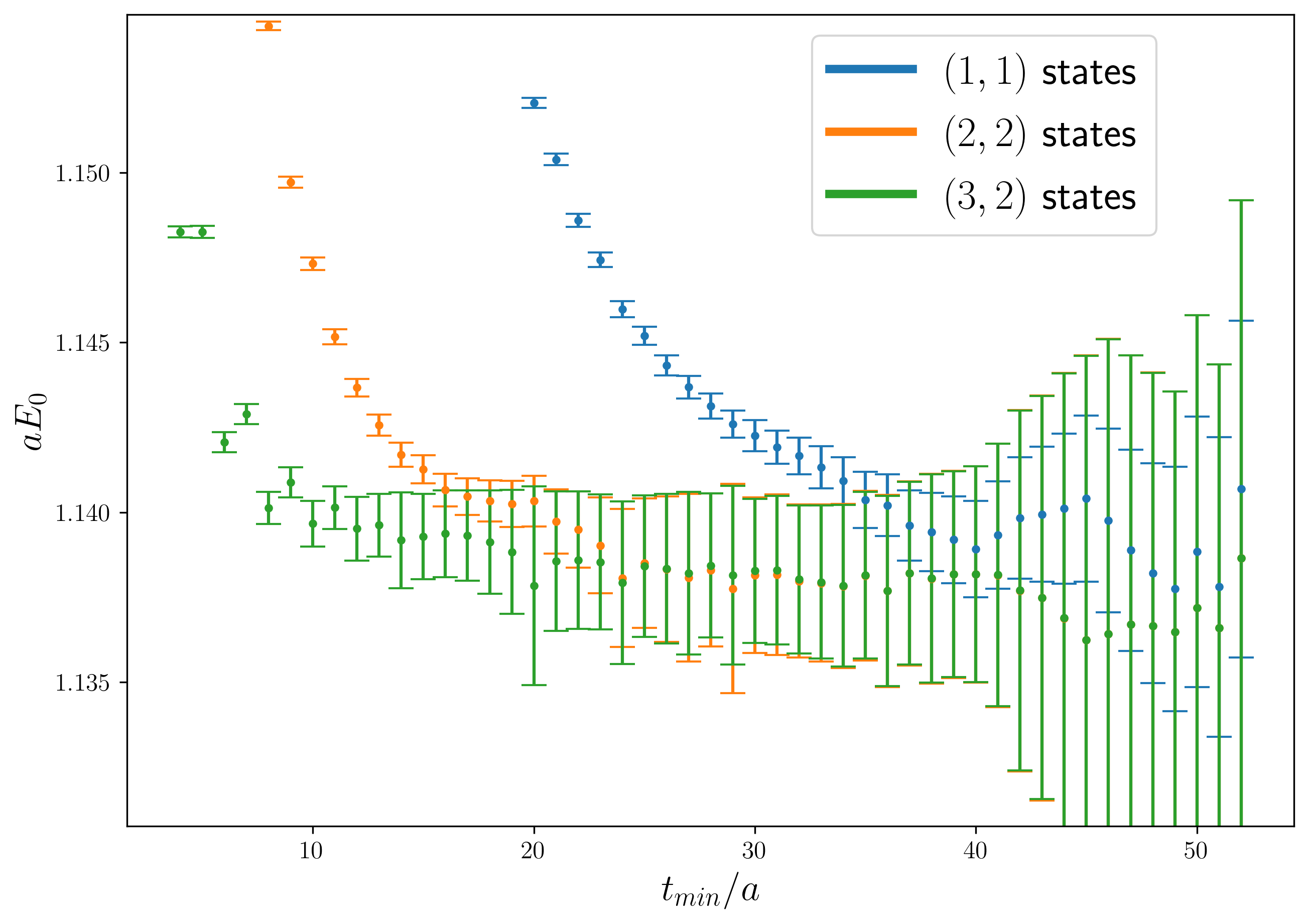}
    \caption{Fit posteriors for the ground-state $B$-meson energy with $am_h \approx am_b$ on the physical-mass ensemble with $a\approx 0.04 \, \rm{fm}$.
    The color indicates the number of states included in the fit to the truncated spectral decomposition.
    The $x$-axis indicates the starting time of the fit.  
    For data points with the same color (i.e. set number of states), the posteriors can be seen to reach a plateau value as $\tmin$ is increased before the signal eventually begins to degrade.
    As more states are included in the fit, stable extractions of the ground-state mass become possible at earlier Euclidean times.
    }
    \label{fig:bmeson-stability}
\end{figure}
To ensure that results are stable with respect to reasonable variations in the fits, the present analysis varies the starting time $\tmin$ of the fit as well as the number of states included in the spectral decomposition.
\Cref{fig:bmeson-stability} shows an example of the stability observed in fits to a heavy-meson two-point function with $M_H \approx 5300$~MeV on the physical-mass $a\approx 0.04$~fm ensemble.
% As an example, see \cref{fig:bmeson-stability}.
The analysis of two-point functions serves to fix the values for $t_{\rm min}$ and the number of states to be used later in joint fits to the two-point and three-point functions.
The posterior values for the ground-state energy and amplitudes are also used as inputs to those later fits. In general terms, the procedure matches that described in \cite{FermilabLattice:2022gku}.

\subsection{Three-point functions}

Suitable ratios of two-point and three-point functions give quantities that plateau to the desired form factors as the source-sink separation becomes large:
\begin{align}
R_{\parallel}(t, T, \bm{p}) &=
     \frac{ \bar{C}_{H \to \finalstate}^{V^0}(t,T, \bm{p}) \sqrt{2 E_\finalstate}}{ \sqrt{\bar{C}_{\finalstate}^{A^0}(t, \bm{p}) \bar{C}_{H}^P(T-t) e^{-E_\finalstate t} e^{-M_H (T-t)}}},
    \label{eq:ratio_v4}\\
R_{\perp}(t, T, \bm{p})     &= \frac{\sqrt{2 E_\finalstate}}{p^i}
    \frac{ \bar{C}_{H \to \finalstate}^{V^i}(t,T, \bm{p})}{ \sqrt{\bar{C}_{\finalstate}^{P}(t, \bm{p}) \bar{C}_{H}^P(T-t) e^{-E_\finalstate t} e^{-M_H (T-t)}}},
    \label{eq:ratio_vi}\\
R_0(t, T, \bm{p}) &= 2 \sqrt{M_H E_\finalstate} \left(\frac{m_h - m_l}{M_H^2 - M_\finalstate^2}\right)
    \frac{ \bar{C}_{H \to \finalstate}^S(t,T, \bm{p})}{ \sqrt{\bar{C}_{\finalstate}^P(t, \bm{p}) \bar{C}_{H}^P(T-t) e^{-E_\finalstate t} e^{-M_H (T-t)}}},
    \label{eq:ratio_s}
\end{align}
where the bars (e.g., $\bar{C}^P_{\finalstate}$) denote the time-slice-averaged versions of the correlators (see \cite{FermilabLattice:2022gku} for additional details and the definition of time-slice averaging).
Up to discretization effects and renormalization, these ratios asymptotically approach the form factors at large Euclidean times:
\begin{align}
    R_\parallel(t, T, \bm{p}) &\stackrel{0 \ll t \ll T}{\longrightarrow} Z_{V^0}^{-1} f_\parallel(\bm{p}),\\
    R_\perp(t, T, \bm{p})     &\stackrel{0 \ll t \ll T}{\longrightarrow} Z_{V^i}^{-1} f_\perp(\bm{p}),\\
    R_0(t, T, \bm{p})         &\stackrel{0 \ll t \ll T}{\longrightarrow} f_0(\bm{p}).
\end{align}
Since the ratios also plateau quite slowly as $T\to \infty$, the present analysis controls for excited states by carrying out fits directly to the spectral decomposition in \cref{eq:2pt_spectral_decomp_final,eq:2pt_spectral_decomp_initial,eq:3pt_spectral_decomp}.
Nevertheless, the ratios continue to provide visual guidance on the presence and size of excited-state effects, which aids in the interpretation of subsequent fit results.
\begin{figure}
    \centering
    \includegraphics[width=0.95\linewidth]{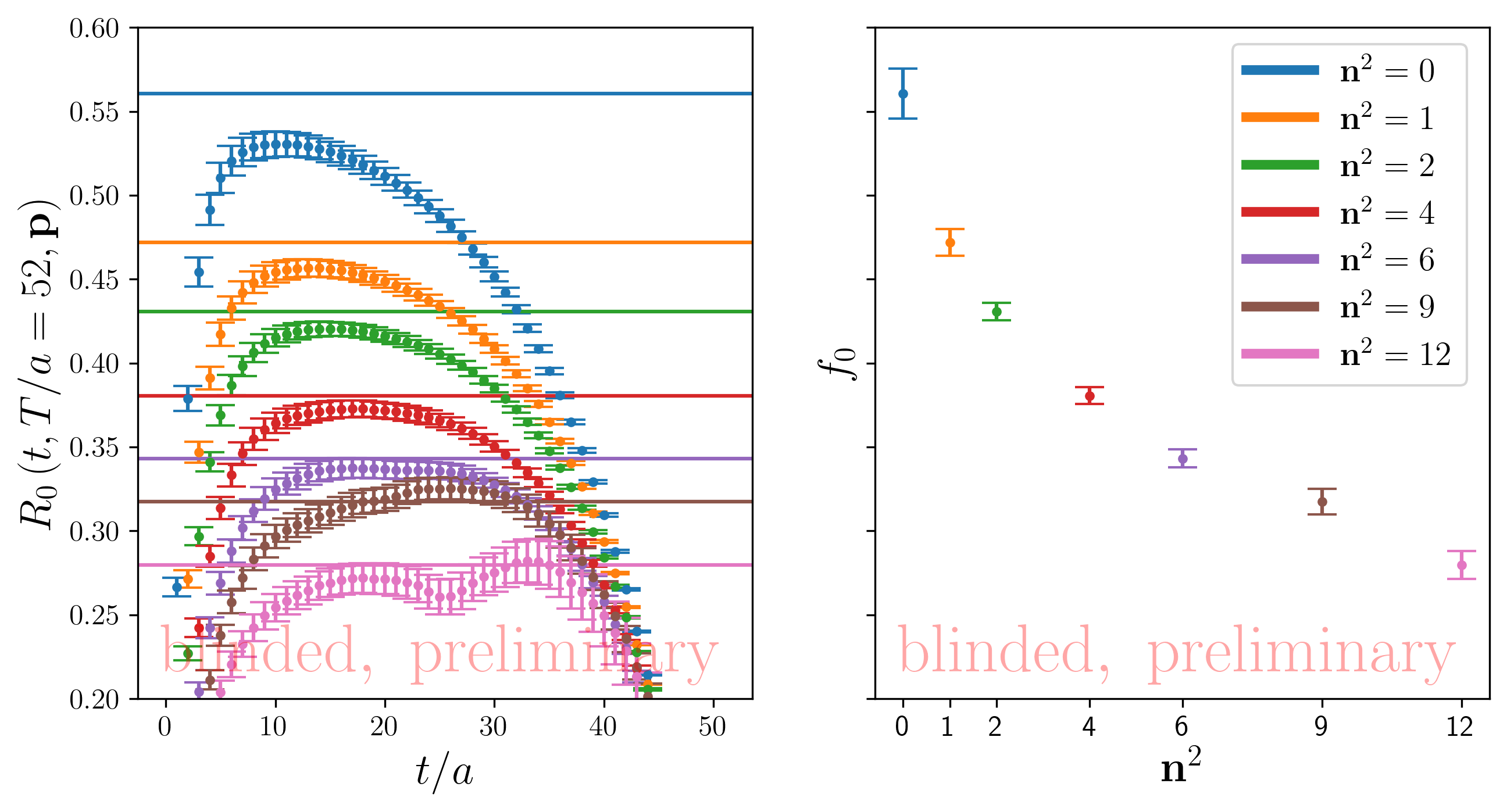}
    \caption{Ratio of two-point and three-point functions for different momenta, compared to fit posteriors for the bare form factors, at $a m_h \approx am_b$ on the physical-mass $a\approx 0.04$~fm ensemble for $B\to\pi$.
    The right panel shows the form factors as a function of the squared momentum of the pion, where $\bm{p}^2 = (2\pi/L)^2 \bm{n}^2$ with $\bm{n} \in \mathbb{Z}^3$. }
    \label{fig:ratplot}
\end{figure}
An example of the ratio $R_0$ for different momenta at the largest source-sink separation ($T/a=52$) is shown in the left panel of \cref{fig:ratplot}.  
The left panel of \cref{fig:ratplot} also shows the central value of the form-factor posteriors resulting from joint correlated fits to the associated two-point and three-point functions;
for the heavy two-point function three non-oscillating and two oscillating states were used, while three non-oscillating states were used for the pion.  Pions at nonzero momentum include an additional oscillating state. 

The fit posteriors are seen to lie somewhat above the ratios, indicating the level of excited-state contributions identified by our analysis procedure.
The right panel of \cref{fig:ratplot} plots the posteriors as a function of squared momentum of the pion $\bm{p}^2 = (\tfrac{2\pi}{L})^2 \bm{n}^2$ with $\bm{n} \in \mathbb{Z}^3$.

\begin{figure}
    \centering
    \includegraphics[width=0.8\linewidth]{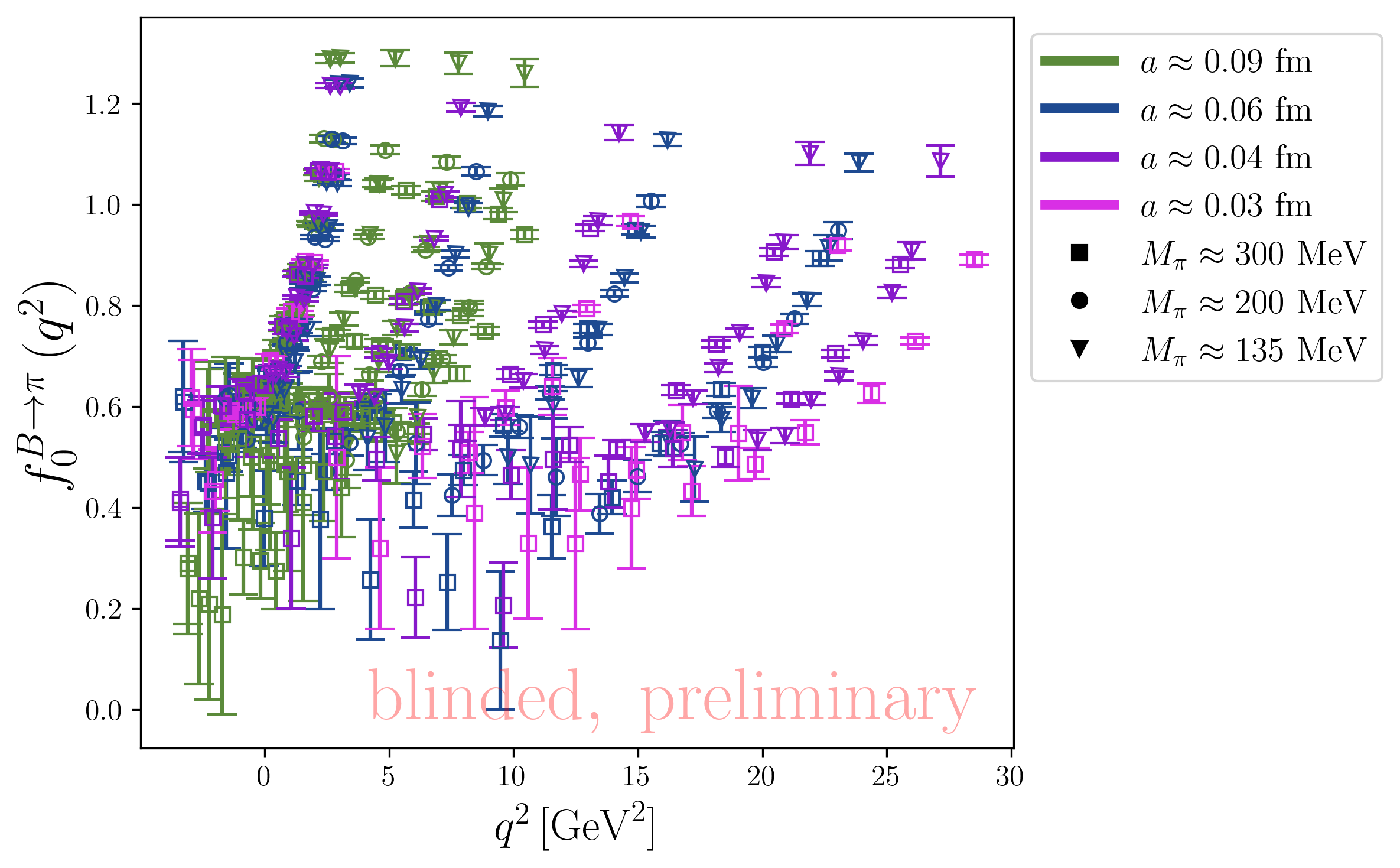}
    \caption{Joint-fit posteriors for scalar form factors across all ensembles, heavy quark masses, and momenta for the decay \Bpi. 
    }
    \label{fig:scalar_ff}
\end{figure}
\begin{figure}
    \centering
    \includegraphics[width = 0.85\linewidth]{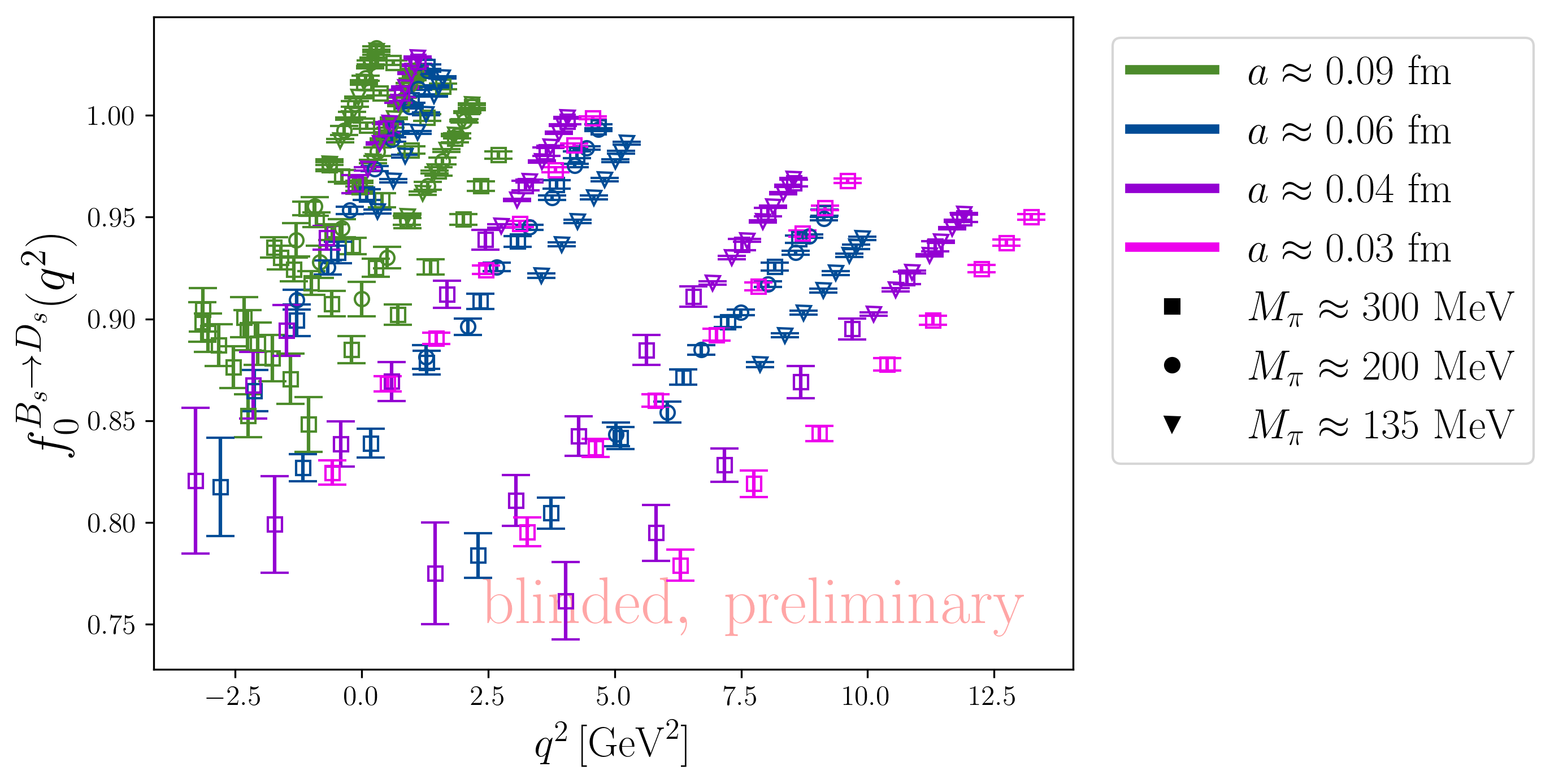}
    \caption{Joint-fit posteriors for scalar form factors across all ensembles, heavy quark masses, and momenta for the decay $B_{s}\to D_{s}$. 
    }
    \label{fig:scalar_ff_bsds}
\end{figure}

\Cref{fig:scalar_ff,fig:scalar_ff_bsds} show fit posteriors for the scalar form factor $f_0$ for the decay $H\to \pi\ell\nu$ and $H_s \to D_s \ell\nu$, respectively, across all ensembles, momenta, and heavy quark masses. Form factors are smoothly rising functions of $q^2$.
Although the data in \cref{fig:scalar_ff} are discrete, this structure is readily apparent.
For instance, on the right-hand side of the figure, the purple or pink sets of points (with a given plotting symbol) trace out a smooth curve.
Data of a fixed color and plotting symbol in \cref{fig:scalar_ff} also show the heavy-quark mass dependence of our results.
For example, the five green circles that appear nearly horizontally in the upper left show the result as the heavy quark mass is increased from $am_h \approx a m_c$ up to $am_h \approx 1$ on the physical-mass $a\approx 0.09$~fm ensemble.
The same behavior is true generically: for a fixed ensemble, form-factor results shift to the right in \cref{fig:scalar_ff} as the quark mass is increased.
Results using the heaviest quarks---with mass values interpolating around the physical $b$ mass---therefore appear on the right of the plot.
By working from $M_H\approx M_D$ to $M_H \approx M_B$, our calculation is able to span essentially the full kinematic range of roughly 25~GeV$^2$, although extrapolation $M_H \to M_B$ remains necessary at low $q^2$. The data shown in \cref{fig:scalar_ff} serve as inputs for the next stage of the calculation, which is the construction of the continuum limit and the evaluation at the physical point. \Cref{fig:scalar_ff_bsds} shows the same for $H_s \to D_s$

\begin{figure}
    \centering
    \includegraphics[width=0.75\linewidth]{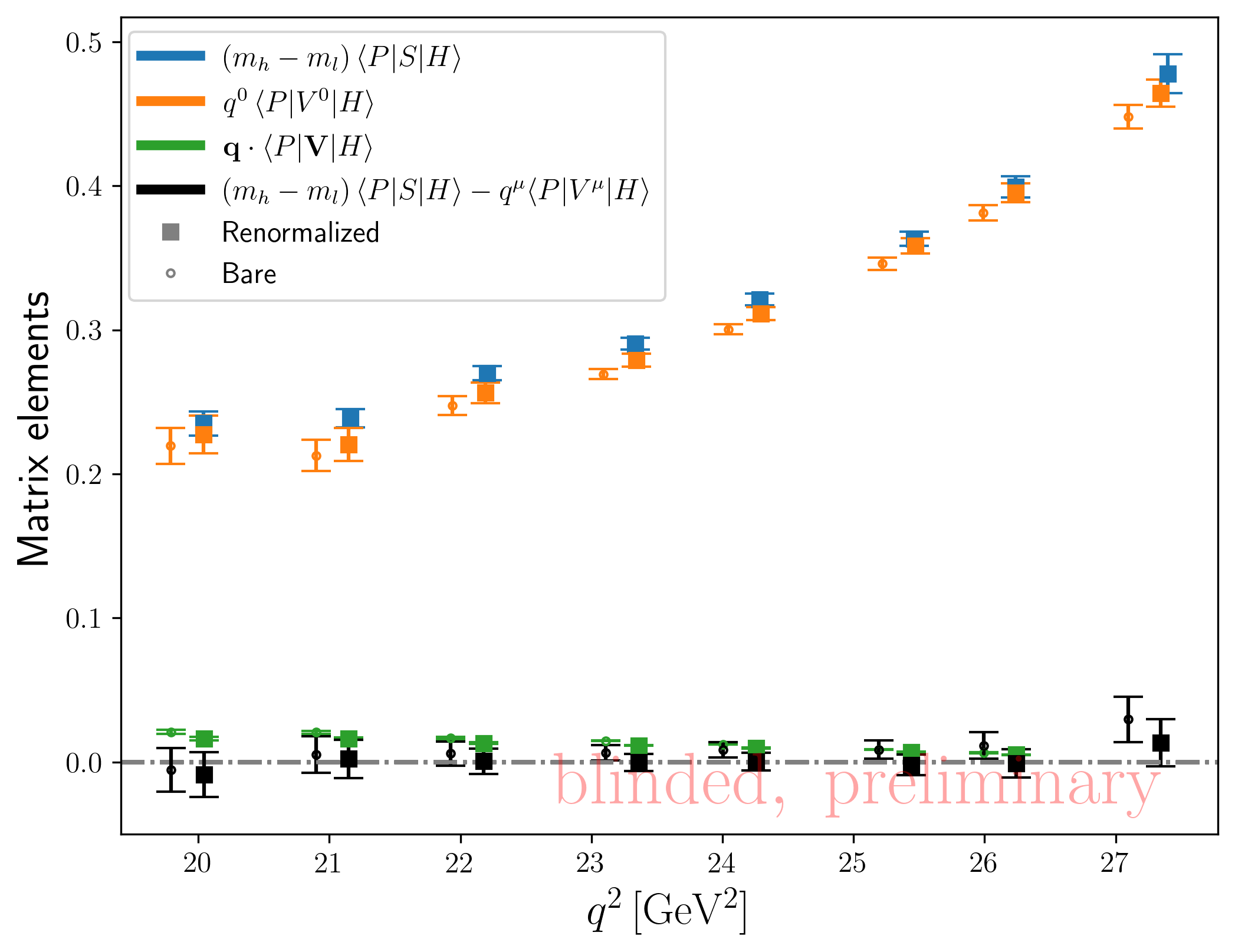}
    \caption{Comparison of bare and renormalized matrix elements from correlators with $am_h\approx a m_b$ on the physical-mass $a\approx0.04 \, \rm{fm}$ ensemble for the decay \Bpi. The colored points indicate data of a single current at different momenta. 
    Non-perturbative renormalization is done by imposing PCVC, \cref{eq:PCVC}, which is shown in black.
    Renormalized values are displayed as closed squares, while open circles are used for bare quantities. For the sake of visual clarity, renormalized quantities have been slightly offset along the x-axis.}
    \label{fig:ward_id}
\end{figure}

So far the discussion has focused on the determination of the bare form factors.
Non-perturbative renormalization is carried out using PCVC, \cref{eq:PCVC}.
Besides furnishing the required renormalization factors $Z_{V^0}$ and $Z_{V^i}$, PCVC provides a useful diagnostic for control of excited-states across different currents and momenta.\footnote{Partial conservation of the axial current has been used in a similar fashion as a diagnostic for the presence of uncontrolled $N\pi$ excited states in calculations of the axial form factor of the nucleon~\cite{Gupta:2017dwj,Gupta:2024krt}.}
\cref{fig:ward_id} compares bare and renormalized matrix elements as a function of $q^2$ near the bottom mass ($am_h \approx am_b$) on the physical-mass $a\approx 0.04$~fm ensemble.
The renormalization factors $Z_{V^0}$ and $Z_{V^i}$ were extracted using a fit to \cref{eq:PCVC}.
Near the bottom mass, PCVC is dominated by contributions from the scalar current and temporal component of the vector current, which scale like $m_b$ and $M_B \approx  m_b + \bar{\Lambda} + \mathcal{O}(1/m_b)$ in heavy-quark effective theory (HQET).

\section{Chiral-continuum fits}
\label{sec:continuum_fits}

The following section focuses on $B_{(s)} \to D_{(s)}$, for which the present analysis is more advanced due to the statistical advantage of having a heavy quark in the initial and final state. 
Chiral continuum results for $B \to \pi$ are in progress. We construct the continuum limit using a fit model based on hard SU(2) $\chi$PT, including chiral logarithms through NLO and analytic terms consistent with the assumed power counting through NNLO.
Building on previous work on $D$-meson semileptonic decays~\cite{FermilabLattice:2022gku}, we model the form factors with an expression of the form
\begin{equation}
\begin{split}
    w_0^{d_P} f_{P}(E)
    = \frac{c_0}{w_0\left(E + \Delta_{xy, P}\right)}
    \times \Big[
    1 &+ \delta f_{\rm logs} + c_l \chi_l + c_s \chi_s + c_C \chi_C +c_H \chi_H + c_E \chi_E \\
    &+ c_{l^2} \chi_l^2 + c_{ls} \chi_l \chi_s + c_{s^2} \chi_s^2\\
    &+ c_{lH} \chi_l \chi_H + c_{lE} \chi_l \chi_E + c_{sH} \chi_s \chi_H + c_{sE} \chi_s \chi_E\\
    & + c_{H^2} \chi_H^2 + c_{HE} \chi_H \chi_E + c_{E^2} \chi_E^2\\
    & + c_{C^2}\chi_C^2 +  c_{CE}\chi_C\chi_E + c_{CH}\chi_C\chi_H + c_{lC}\chi_l\chi_C\\
    & + \chi_{sC}\chi_s\chi_C
    + \delta f_{\rm artifacts}^{(a^2 + h^2+c^2)}
    \Big],
\end{split}
\label{eq:chipt_hard_su2}
\end{equation}
where the analytic terms $\chi$ take into account energy dependence, mass mistunings, and heavy -quark mass dependence. With multiple charm final states, we are able to isolate the charm discretization effects as well as interpolate to the $D_{(s)}$. 
Explicit forms for the analytic terms and chiral logarithms are given in \cite{FermilabLattice:2022gku}.
The location of the pole is given by
$\Delta_{cs,P} = (M_{H_{c}^*(J^P)}^2-M_{H_{s}}^2-M_{D_{s}}^2)/2M_{H_{s}},$
where pole masses are taken from \cite{Dowdall:2012ab}. For intermediate masses, we interpolate between $\chi_c(J/\psi)$ and $B_{c0}(B_c^*)$ for the scalar and vector form factors.  

\begin{figure}
    \centering
    \includegraphics[width=0.95\linewidth]{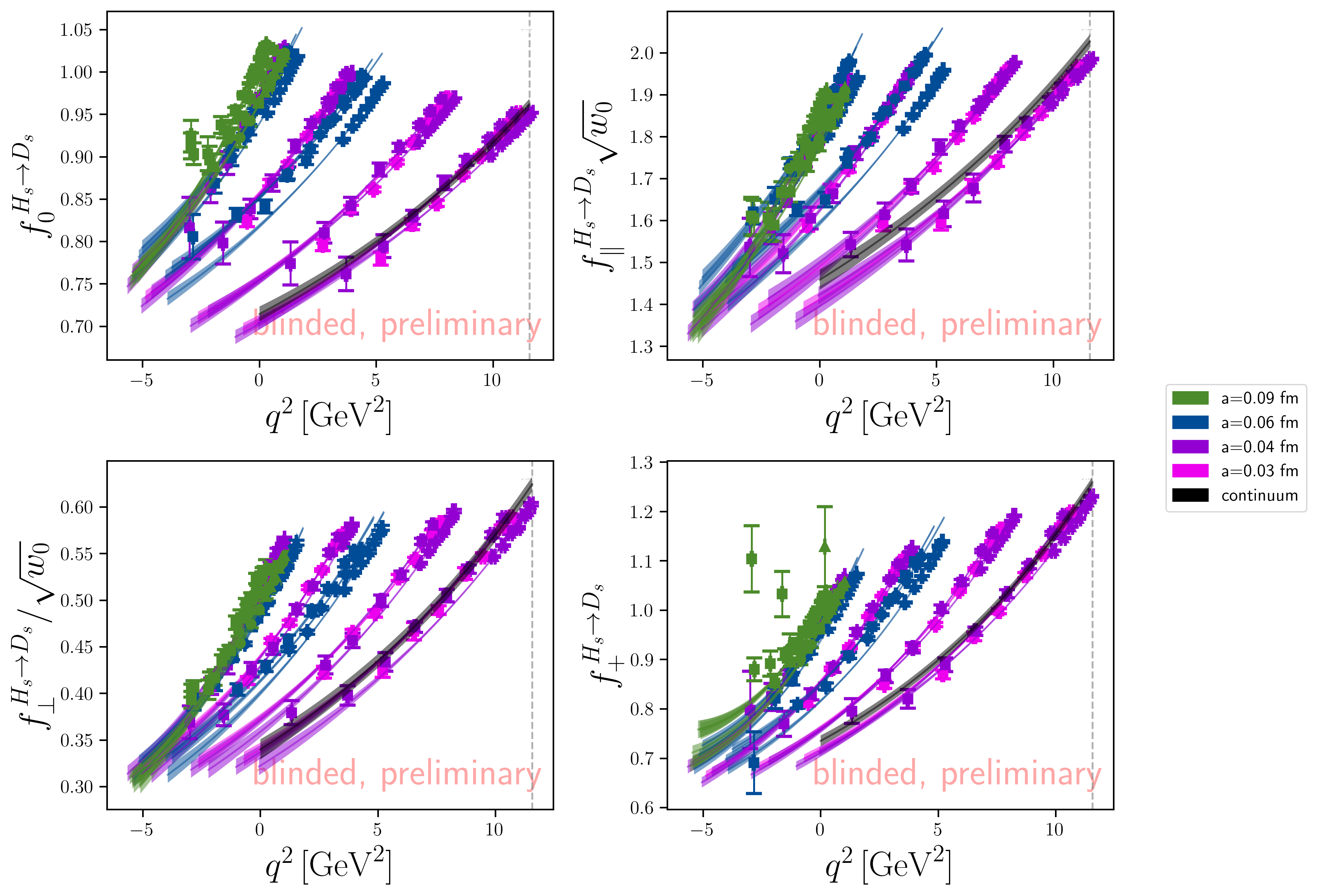}
    \caption{Chiral continuum-limit fits  for the form factors $f_0(q^2), f_\parallel(q^2), f_\perp(q^2)$, and $f_+(q^2)$ for the decay $B_s \to D_s$.
    Results are shows in units of $w_0$.
    The black curve denotes the continuum limit at the physical point.
    }
    \label{fig:continuum_limit}
\end{figure}
Figure \ref{fig:continuum_limit} shows the continuum limit for the renormalized form factors in units of $w_0$. We construct $f_+$ through \cref{eq:fplus_perp_0}. As the colored bands show visually, the resulting fit describes the input data well on each ensemble.
Although our analysis of statistical and systematic uncertainties continues, the preliminary fits shown in \cref{fig:continuum_limit} all satisfy $\chi^2/{\rm DOF} \simeq 1$. We note larger errors on the 0.09~fm 0.2$m_s$ ensemble which is due to correlations that affect all decay channels. 

\begin{figure}
    \centering
    \includegraphics[width=\linewidth]{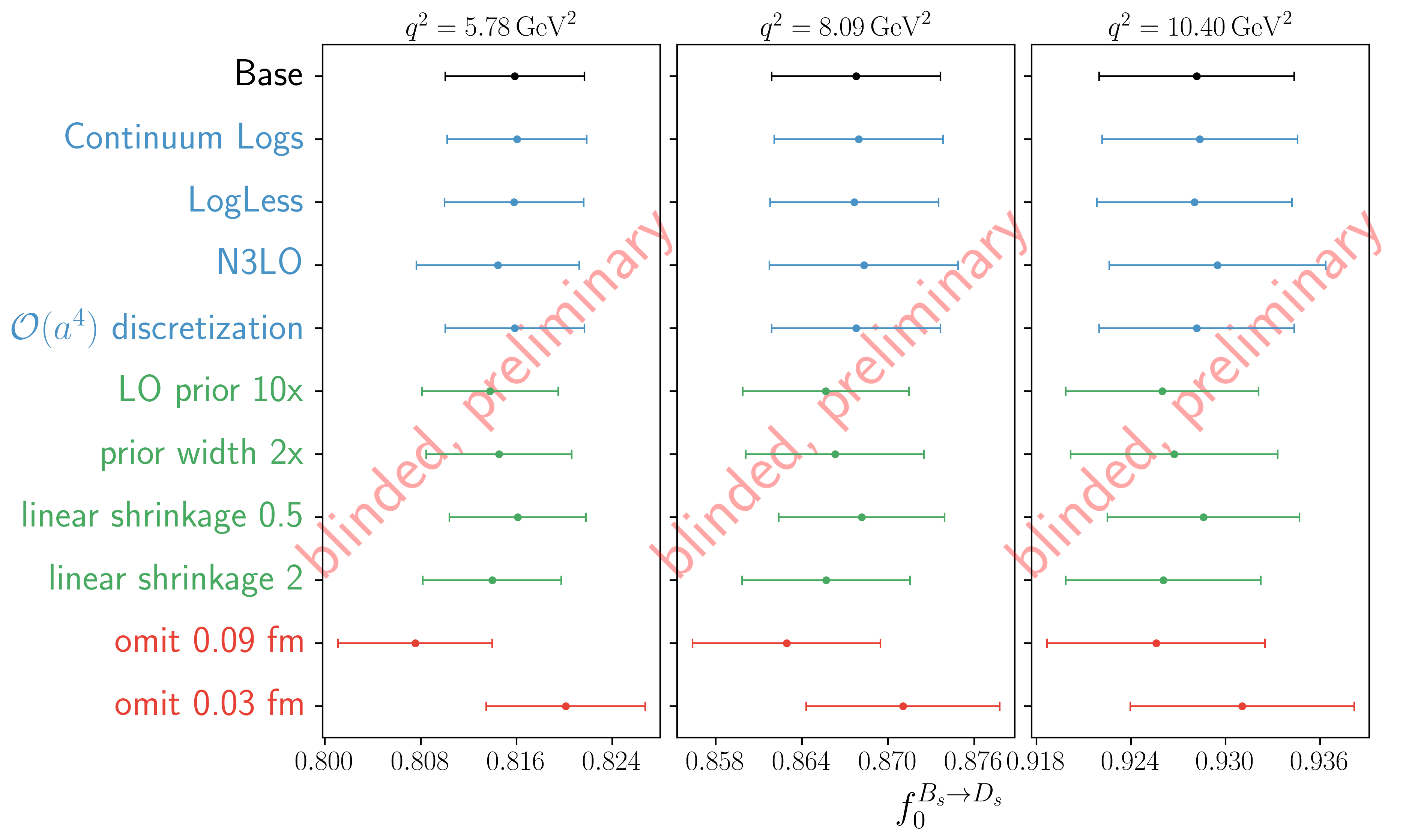}
    \caption{Stability of $f_0^{B_s \to D_s}$ under variations of the fit function for $q^2$ values that feed into the $z$~expansion.
    }
    \label{fig:stability}
\end{figure}
Figure \ref{fig:stability} shows the stability of $f_0$ for $B_s\to D_s$ at three different kinematic points within the data region under different variations of the continuum fitting procedure. 
The base fit function is defined in \cref{eq:chipt_hard_su2}.
The first class of variations shown in blue constitute modifications to \cref{eq:chipt_hard_su2}, varying the structure of the chiral logarithms and adding higher-order analytic terms or discretization effects.
The second class of variations are modifications to the fitting itself, demonstrating stability under reasonable expansion of the priors and under mild adjustment of the shrinkage applied to the covariance matrix.
Finally, the last pair of variations checks for stability upon dropping the coarsest or finest ensemble.
All variations are consistent to much better than one standard deviation; in fact, apart from dropping the coarsest and finest ensembles, the results are essentially indistinguishable.
Overall, this stability demonstrates that the continuum results are insensitive to choices made during the analysis of form factors. Corresponding figures for $B \to D$ show similar behavior and will be presented in detail in a upcoming publication.

\section{\texorpdfstring{\boldmath$z$}{z} expansion}
\label{sec:z-expansion}

\begin{figure}
    \centering
    \includegraphics[width=0.49\linewidth]{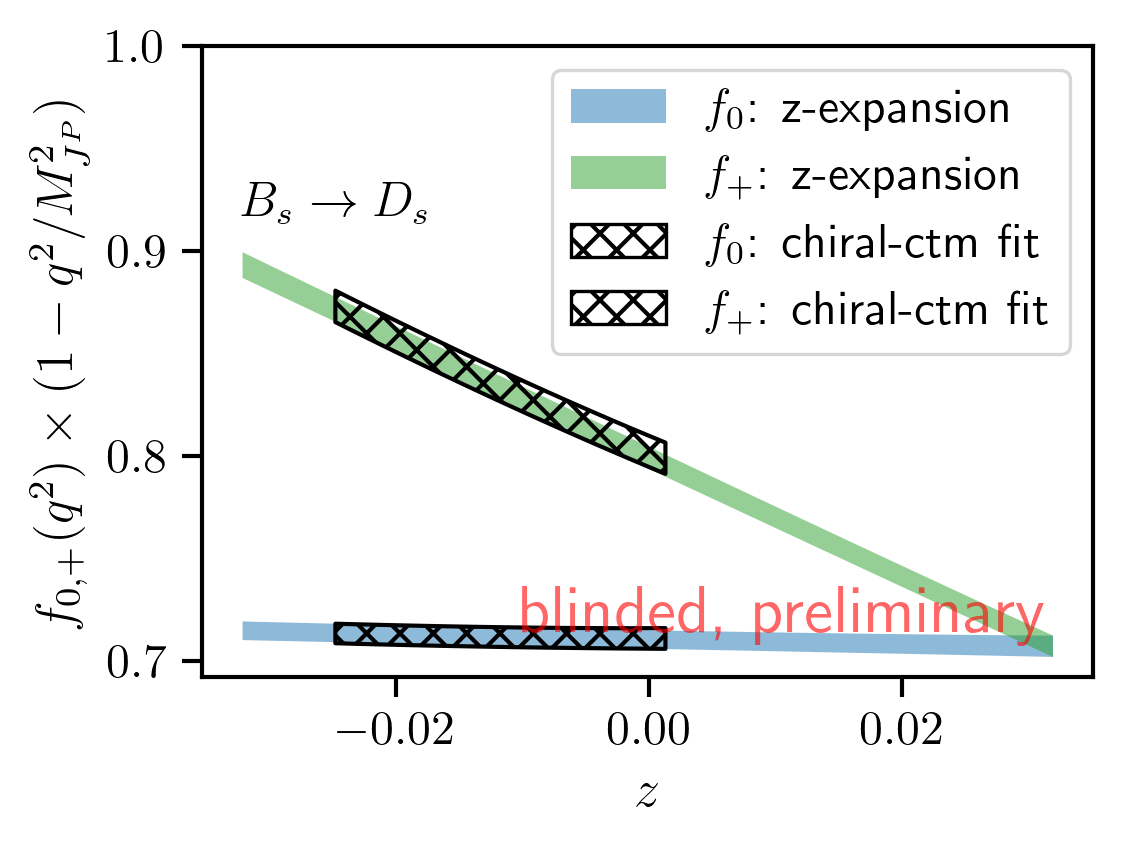}
    \includegraphics[width =0.49\linewidth]{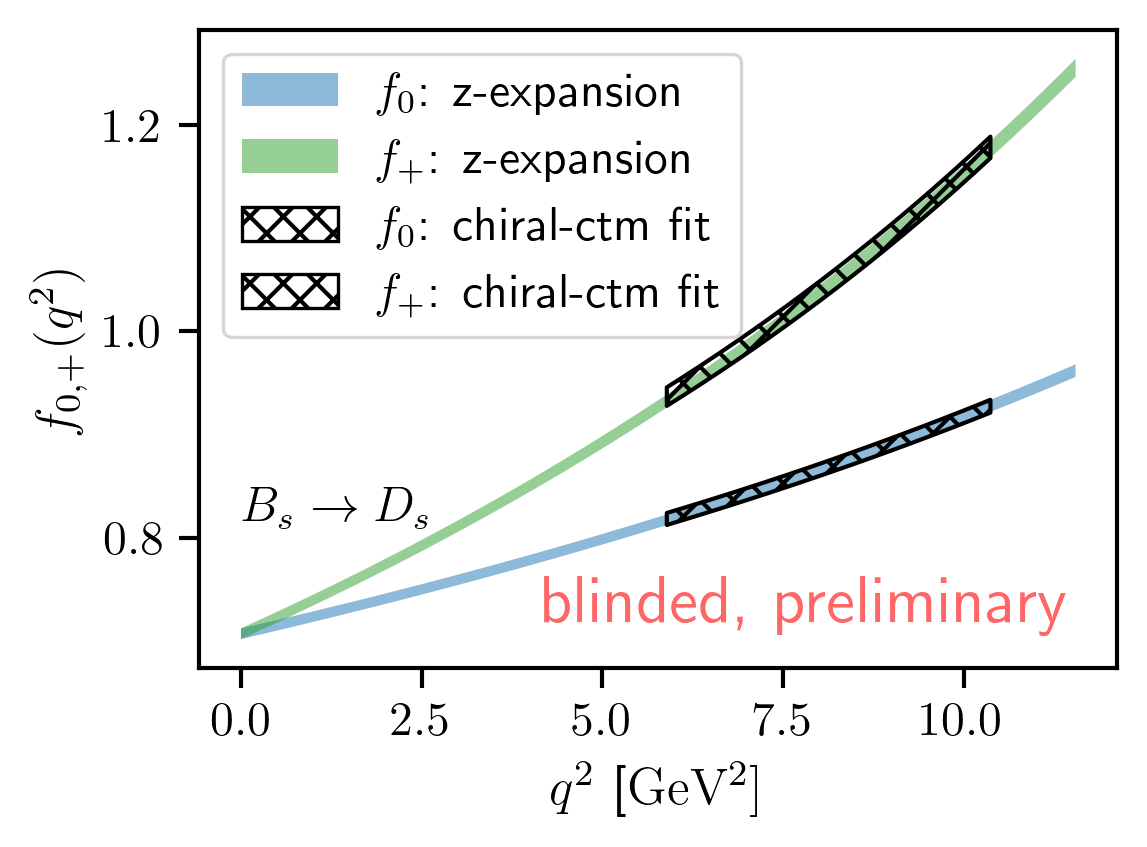}
    \caption{Comparison of different parameterizations of the continuum form factors.
    The solid bands show the $z$~expansion, while the hashed regions use the model from \cref{eq:chipt_hard_su2}.
    The hashed regions span the kinematic range of the synthetic data used in the fits to the $z$~expansion.
    \label{fig:z-expansion}
    }
\end{figure}

The continuum limit constructed above is expected to be reliable in the kinematic region spanned by our lattice QCD calculation.
Our calculation employs both physical-mass and lighter-than-physical $b$ quarks and is able to achieve good kinematic coverage even down to $q^2\approx0$.
To compare with experimental results, it is convenient to reparameterize the form factors using the so-called $z$~expansion, which exploits their analytic structure.
Consider the change of variables defined by the conformal transformation
\begin{align}
    z(q^{2}, t_{0}) = \frac{\sqrt{t_{+} - q^{2}} - \sqrt{t_{+} - t_{0}}}{\sqrt{t_{+} - q^{2}} + \sqrt{t_{+} - t_{0}}},
\end{align}
where $t_+ = (M_{D}+M_{B})^2$ is the start of the multiparticle cut and $t_0$ can be chosen to optimize the mapping.
This change of variables maps the complex-$q^2$ plane to the unit disk allowing for the form factors to be expressed as a hyperconvergent series of $z$.
Several different parameterization for the form factors have been used in the literature on semileptonic decays;
they differ primarily in the choice of additional theoretical assumptions about the structure form factors (e.g., knowledge from perturbative QCD). We explore both the BGL and BCL parameterizations \cite{Boyd_1995,Bourrely:2008za}. The present analysis focuses on the BCL parameterization~\cite{Bourrely:2008za}:
\begin{align}
f_0(z) &= \frac{1}{1 - q^2(z)/M_{0+}^2}
\sum_{m=0}^{M-1} b_m z^m , \\
f_+(z) &= \frac{1}{1 - q^2(z)/M_{1-}^2}
\sum_{n=0}^{N-1} a_n \left( z^n - \frac{n}{N} (-1)^{\,n-N} z^N \right) .
\end{align}
Here we focus on our results for $B_s \to D_s$ Our preferred fits use $N = M = 3$, using three synthetic data points in the data region as inputs. The kinematic constraint $f_0(0) = f_+(0)$ is imposed. We choose an optimal $t_0 = t_+ - \sqrt{t_+(t_+-t_-)}$ where $t_- = (M_{B_s}-M_{D_s})$ that symmetrizes the values of $z$ because our lattice data do not span the entire kinematic region.
\Cref{fig:z-expansion} compares the result of the continuum extrapolation from the $\chi$PT fits with that of the $z$~expansion, demonstrating essentially identical central values for the two parameterizations. 

\begin{figure}
    \centering
    \includegraphics[width=0.99\linewidth]{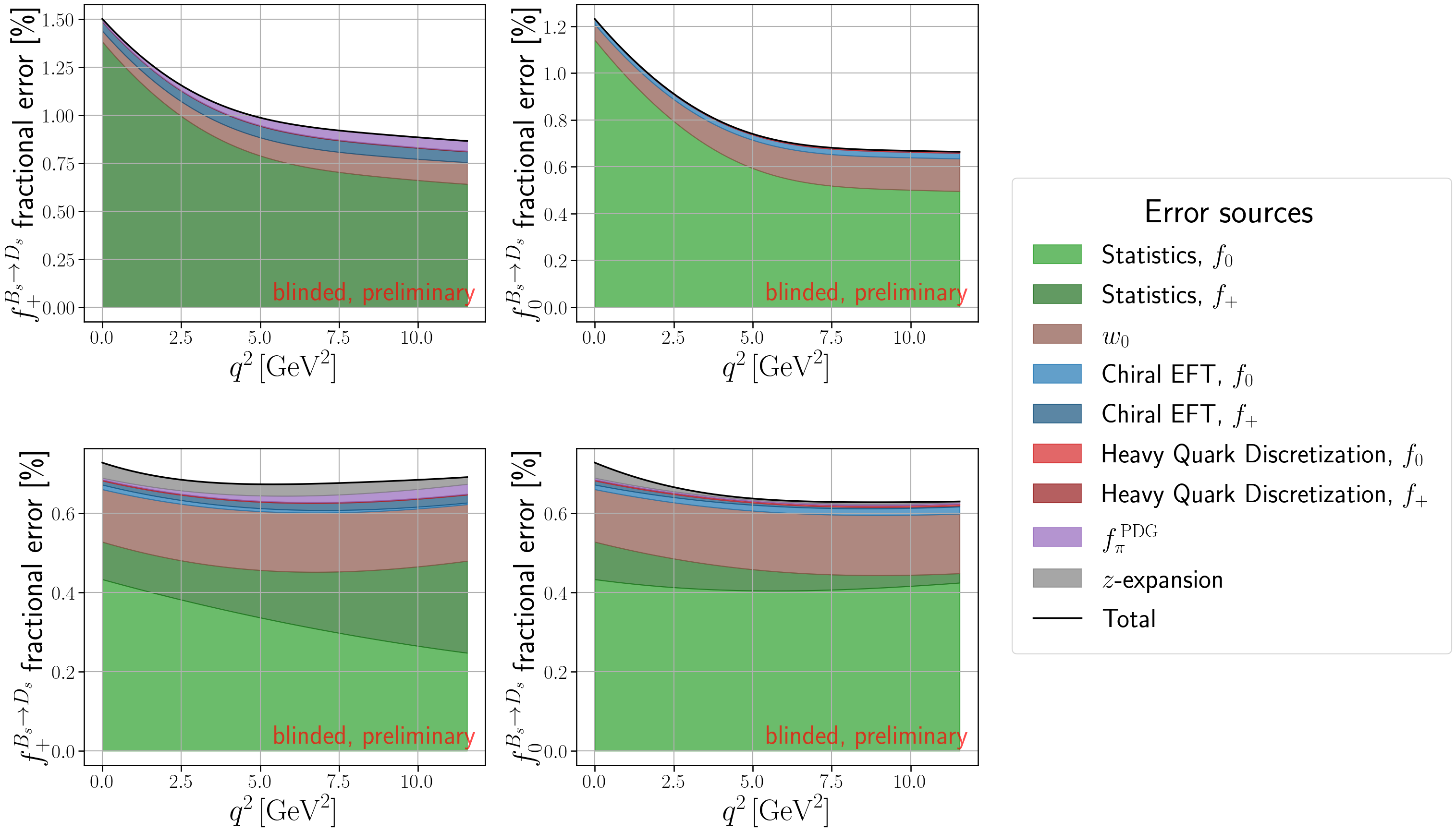}
    \caption{
    Comparison of the preliminary error budgets for the continuum form factors using the $z$~expansion and the parameterization of \cref{eq:chipt_hard_su2}. The top row is the error budget from coming from the $\chi$PT continuum fits while the second row is the error budget coming from the $z$-expansion fits. 
    \label{fig:error_budget}
    }
\end{figure}
\Cref{fig:error_budget} shows a preliminary error budget for the continuum form factors across the kinematic range. In line with previous experience~\cite{PhysRevD.85.114502}, the rapidly convergent parameterization of the $z$~expansion yields a decrease in overall uncertainties.
The dominant uncertainty is the statistical uncertainty associated with the underlying form-factor fits on each ensemble.
We expect to decrease this uncertainty through additional running, particularly on the finest ensembles. For $B_s \to D_s$ we can see that the overall uncertainty is less than $1\%$ over the kinematic range. Preliminary total uncertainties are already approaching our overall precision goals. 

\section{Conclusion}
We have provided an update of ongoing work by the Fermilab Lattice and MILC collaborations to calculate hadronic form factors for semileptonic decays of $B_{(s)}$ mesons using the HISQ action for all sea and valence quarks.
A distinguishing feature of this calculation is the inclusion of an ensemble with $a\approx 0.04$~fm and physical light quark masses. \Cref{sec:correlator_analysis} discussed the common fitting strategy, used for all decay channels, to extract bare lattice form factors from joint correlated fits to two-point and three-point functions.
\Cref{sec:continuum_fits,sec:z-expansion} discussed the preliminary results for heavy-to-heavy form factors which were presented at the conference.
Similar fits are also underway for $B\to\pi\ell\nu$.

In summary, the current results show progress toward our eventual goal of percent-level precision to match that of expected near-term experimental gains, e.g., at Belle II and LHCb.
The present work is a part of our collaboration's longer term goal of providing Standard Model predictions for tree-level and rare semileptonic decays of $B$-mesons to final states with a single pseudoscalar hadron.
Future plans include a study of $B_s \to K$ together with a correlated analysis to extract the ratio $\Vub/\Vcb$ from $\mathcal{B}{(B_s \to K\ell \nu)}/\mathcal{B}(B_s \to D_s\ell\nu)$ as well as a study of the rare decay $B \to K\ell\nu$.

\section*{Acknowledgments}
%%NERSC
This research used resources (especially Franklin and Perlmutter) of the National Energy Research Scientific Computing Center (NERSC), a U.S.\ Department of Energy Office of Science User Facility located at Lawrence Berkeley National Laboratory, operated under Contract No.\ DE-AC02-05CH11231.
%% DOE/INCITE/ALCC
We also used Mira, Theta, and Polaris at the Argonne Leadership Computing Facility (ALCF) and Summit, Crusher, and Frontier at the Oak Ridge Leadership Computing Facility (OLCF) under the Innovative and Novel Computational Impact on Theory and Experiment (INCITE)
and ASCR Leadership Computing Challenge (ALCC) programs.
The ALCF and OLCF are DOE Office of Science User Facilities supported under contract Nos.\ DE-AC02-06CH11357 and DE-AC05-00OR22725, respectively.
%%USQCD
Computations for this work were carried out in part with computing and long-term storage resources provided by the USQCD Collaboration.
%%XSEDE ACCESS   Ranch , Frontera
This work used the Extreme Science and Engineering Discovery Environment (XSEDE) storage system Ranch at the Texas Advanced Computing Center (TACC) through allocation TG-MCA93S002.
The XSEDE program was supported by the National Science Foundation under grant No.\ ACI-1548562.
This work also used Ranch at TACC through allocation MCA93S002 from the Advanced Cyberinfrastructure Coordination Ecosystem: Services \& Support (ACCESS) program, which is supported by U.S.\ National Science Foundation grants Nos.\ 2138259, 2138286, 2138307, 2137603, and 2138296.
Also through ACCESS allocation MCA93S002 this research used both the DeltaAI advanced computing and data resource, which is supported by the National Science Foundation (award OAC 2320345) and the State of Illinois, and the Delta advanced computing and data resource which is supported by the National Science Foundation (award OAC 2005572) and the State of Illinois. Delta and DeltaAI are joint efforts of the University of Illinois Urbana-Champaign and its National Center for Supercomputing Applications.
This research is part of the Frontera computing project at the Texas Advanced Computing Center.
Frontera is made possible by National Science Foundation award OAC-1818253.
%%Big Red II+ , etc.
Computations on the Big Red II+, Big Red 3, Quartz, and Big Red 200 computers were supported in part by Lilly Endowment, Inc., through its support for the Indiana University Pervasive Technology Institute.
The parallel file system employed by Big Red II+ was supported by the National Science Foundation under Grant No.~CNS-0521433.
%%Blue Waters
Some of the computations were done using the Blue Waters sustained-petascale computer, which was supported by the National Science Foundation (awards OCI-0725070 and ACI-1238993) and the state of Illinois.
Blue Waters was a joint effort of the University of Illinois at Urbana-Champaign and its National Center for Supercomputing Applications (NCSA).

This work was in part based on the MILC collaboration's public lattice gauge theory code~\cite{MILCcode} with QUDA~\cite{Clark_2010,Babich_2011} used to accelerate quark propagator solves and smearing on~GPUs.

This work was supported in part by the U.S.~Department of Energy, Office of Science, under Awards
No.~DE-SC0010120 (S.G.),
No.~DE-SC0015655 (A.X.K., A.C., A.T.L.),
No.~DE-SC0009998 (J.L.),
and “High Energy Physics Computing Traineeship for Lattice Gauge Theory” No. DESC0024053 (A.C.);
by the National Science Foundation under Grants Nos.~PHY20-13064 and PHY23-10571 (C.D., A.V.);
by the Simons Foundation under their Simons Fellows in Theoretical Physics program (A.X.K.);
by MICIU/AEI/10.13039/501100011033 and FEDER (EU) under Grant PID2022-140440NB-C21 (E.G.);
by Consejeria de Universidad, Investigaci\'on y Innovaci\'on and Gobierno de Espa\~na and EU--NextGenerationEU, under Grant AST22~8.4 (E.G.);
by AEI (Spain) under Grant No.\ RYC2020-030244-I / AEI / 10.13039/501100011033 (A.V.);
%%% visits, workshops, programs
A.X.K.\ is grateful to the Pauli Center for Theoretical Studies and the ETH Z\"urich for support and hospitality.
A.X.K. and A.S.K. are grateful to the Kavli Institute for Theoretical Physics (KITP) for hospitality and support during the program ``What is Particle Theory?''
The KITP is supported in part by the National Science Foundation under Grant No.\ PHY-2309135.
This document was prepared by the Fermilab Lattice and MILC Collaborations using the resources of the Fermi National Accelerator Laboratory (Fermilab), a U.S.\ Department of Energy, Office of Science, HEP User Facility.
Fermilab is managed by Fermi Forward Discovery Group, LLC, acting under Contract No.~89243024CSC000002 with the U.S.\ Department of Energy.

\bibliographystyle{unsrt}
\bibliography{refs}

\end{document}